\def\f{s_{f}}
\def\i{s_{i}}
\def\n{s_{n}}
\def\N{s_{n+1}}
\def\R{\rightarrow}
\def\d{\delta}
\def\S{\Sigma}
\def\k{\frac{dx^a}{ds}}
\def\p{\partial}

\documentclass[prd,aps,eqsecnum,nofootinbib,amsmath,amssymb,tightenlines,superscriptaddress]{revtex4}

\usepackage{amsthm,amsfonts,graphicx,verbatim,color,cancel,amsmath}

\makeatletter
\newcommand*\rel@kern[1]{\kern#1\dimexpr\macc@kerna}
\newcommand*\widebar[1]{%
  \begingroup
  \def\mathaccent##1##2{%
    \rel@kern{0.8}%
    \overline{\rel@kern{-0.8}\macc@nucleus\rel@kern{0.2}}%
    \rel@kern{-0.2}%
  }%
  \macc@depth\@ne
  \let\math@bgroup\@empty \let\math@egroup\macc@set@skewchar
  \mathsurround\z@ \frozen@everymath{\mathgroup\macc@group\relax}%
  \macc@set@skewchar\relax
  \let\mathaccentV\macc@nested@a
  \macc@nested@a\relax111{#1}%
  \endgroup
}
\makeatother
\newcommand{\ea}{\end{align}}
\newcommand{\bea}{\begin{align}}

\newcommand{\rn}{\rho_{A}(\N)}
\newcommand{\la}{\langle}
\newcommand{\ra}{\rangle}

\newcommand{\Tr}{{\rm Tr}}

\usepackage{bm}%	bold math
\usepackage{amsfonts}
\usepackage{amssymb}
\usepackage{array}
\usepackage{amsmath}
\numberwithin{equation}{section}
\numberwithin{figure}{section}
\begin{document}
\title{Entanglement Temperature and Entanglement Entropy of Excited States }
%\author{Gabriel Wong${}^1$, Israel Klich${}^1$, Leopoldo A. Pando Zayas${}^2$  and Diana Vaman${}^1$}

\author{Gabriel Wong}
\affiliation
    {%
    Department of Physics,
    University of Virginia, Box 400714,
    Charlottesville, Virginia 22904, USA
    }%
\author{Israel Klich}
 \affiliation
    {%
    Department of Physics,
    University of Virginia, Box 400714,
    Charlottesville, Virginia 22904, USA
    }%
    
    \author{Leopoldo A. Pando Zayas}
    \affiliation{Michigan Center for Theoretical Physics,
    The University of Michigan, 
    Ann Arbor, MI 48109, USA}

\author{Diana Vaman}
 \affiliation
    {%
    Department of Physics,
    University of Virginia, Box 400714,
    Charlottesville, Virginia 22904, USA
    }%

\date {\today}

\begin {abstract}%
{%
 We derive a general relation between the ground state entanglement Hamiltonian and the physical stress tensor within the path integral formalism.
 For spherical entangling surfaces in a CFT, we reproduce
the \emph{local} ground state entanglement Hamiltonian derived by Casini, Huerta and Myers. The resulting reduced density matrix can be characterized by a spatially varying ``entanglement temperature.''  Using the entanglement Hamiltonian, we calculate the first order change in the entanglement entropy due to changes in conserved charges of the ground state, and find a local first law-like relation for the entanglement entropy. Our approach provides a field theory derivation and generalization of recent results obtained by holographic techniques. However, we note a discrepancy between our field theoretically derived results for the entanglement entropy of excited states with a non-uniform energy density and current holographic results in the literature. Finally, we give a CFT derivation of a set of constraint equations obeyed by the entanglement entropy of excited states in any dimension. Previously, these equations were derived in the context of holography.

  }%
\end {abstract}

\maketitle
\section{ Introduction and Motivation}
 For a given quantum state  of a many-body system with density matrix $\rho$, measurements of observables $O_{A}$ supported inside a spatial subregion $A$ are determined by the reduced density matrix $\rho_{A}$, defined by
\begin{equation}\label{red}
\Tr(\rho O_{A})=\Tr_{A}(\rho_{A}O_{A}).
\end{equation}
The relation above is defined to hold for all operators $O_{A}$ in $A$. It follows that $\rho_{A}=\Tr_{B}(\rho)$, where the trace is taken over the complement $B=A^{c}$ .
Since an observer in $A$ has no direct access to degrees of freedom in $B$, he/she suffers a loss of information that can be quantified by the entanglement entropy:
\begin{equation}\label{ent}
S_{A}=-\Tr_{A}(\rho_{A} \ln \rho_{A}).
\end{equation}
$S_{A}$ provides a measure of the entanglement between $A$ and $B$, since increasing the entanglement between $A$ and $B$ will increase the loss of information upon restriction to $A$.  %For a pure state $\rho$, the absence of entanglement between $A$ and $B$ implies the vanishing of $S_{A}$.  
The study of entanglement entropy was originally motivated by attempts to interpret black hole entropy as information loss by an observer restricted to the outside of the event horizon \cite{bombelli1986quantum}.  More recently, entanglement entropy has become an important tool in condensed matter physics, where it plays a role as a diagnostic of many body states. Indeed, the scaling of entanglement entropy characterizes the amenability of systems to numerical simulations such as the density matrix renormalization algorithm (DMRG) in 1d, and the nature of the challenge in higher dimensions. 
An important class of applications of entanglement entropy studies are topological states. Such states have no local observables which reveal their nature, and thus the entanglement entropy may be used in such a situation where no obvious way exists to identify the topological order. For example in \cite{jiang2012identifying} the interplay of DMRG and entanglement entropy on a torus has been used to identify the nature of topological degeneracy. 

While entanglement entropy provides an important measure of entanglement, the reduced density matrix \(\rho_{A}\) is a more fundamental object.  In particular, the study of the entanglement spectrum, i.e. the eigenvalues of $\rho_{A}$,  has picked up pace as it has been recognized as a tool for probing topological order in a more detailed way. For example, the relation between the entanglement entropy of Quantum Hall wave functions and the edge theory associated with such states has been elaborated in \cite{li2008entanglement,lauchli2010disentangling,chandran2011bulk,qi2011general}. Entanglement spectrum also holds a direct relation to the gluing function as well as the gapless edge modes in topological insulators \cite{turner2010entanglement,fidkowski2011model}. These  remarkable relations between a bulk property and edge physics highlight the wealth of information encoded in the entanglement spectrum.

As stated above, the entanglement spectrum of quantum systems may reveal a lot about their nature.
Even more detailed information is available if one knows the actual eigenstates of $\rho_A$. Since any $\rho_A$ is Hermitean and positive semidefinite, it may be expressed as:
\begin{eqnarray}
\rho_A=e^{-H_{A}}
\end{eqnarray}
for some Hermitean operator $H_{A}$. If $H_{A}$ is known, the detailed study of $\rho_A$ follows immediately to the exploration of $H_{A}$. Unfortunately, in most cases $H_{A}$ does not offer a particular simplification or advantage as it is in general a highly nonlocal operator. 

However, in particular special cases $H_{A}$ may become local and simple enough to be used for calculations. The prime example for such a situation has arisen as a result of studies of Hawking and Unruh radiation. According to the Bisognano-Wichmann theorem  \cite{bisognano1975duality,bisognano1976duality}, the causal evolution of a quantum field theory where $A$ is taken to be a half space may be described by a modular operator which is generated by a Lorentz boost. The Minkowski ground state in a causal wedge is then shown to satisfy a  Kubo-Martin-Schwinger condition with respect to the boost, establishing $H_{A}$ as the generator of Lorentz boost.

Recently this result was extended by  Casini, Huerta and Myers \cite{myers}. For a spherical region $A$ in a CFT, they find that the entanglement Hamiltonian may be written explicitly in a local form using the physical energy density \(T_{00}\):
\begin{equation}
\label{Ha}
 H_{A}=\int_{A}\beta(x)T_{00}(x).
 \end{equation}
In this paper, we use the locality property of entanglement Hamiltonians such as \eqref{Ha} to compute the entanglement entropy of excited states. 

 The starting point of our story is an elementary derivation of the above formula using the representation of the ground state reduced density matrix \( \langle\phi | \rho_{A}| \phi '\rangle \) as a Euclidean Path integral integral with boundary conditions for the fields \(\phi\) and \(\phi '\) along the cut at $A$ \cite{hlw}.

Deferring the explicit derivation to section II, let us first discuss the basic idea. 
Treating \(\rho_{A}\) as a propagator, we  derive the expression (\ref{Ha})  by performing the path integral along a Euclidean "time" $s$ that evolves the upper edge of A to the bottom. %In other words, we are looking for a transfer matrix for \(\rho_{A}\).
The resulting path integral may be expressed as: 
\begin{equation}
\label{Texp}
\rho_{A} =Z^{-1}_{A} T \exp \{- \int_{\i}^{\f} K(s) ds\} ,
 \end{equation}
 % The operator K here may include anomalous contribution from the path integral measure %
 where  $T$ denotes "time" ordering in $s$ and $K$ is the quantum operator generating $s$ evolution.   

If the path integral of our theory is invariant under translations in $s$, then $K$ is a conserved charge independent of ``entanglement time'' $s$.  Hence:
 \begin{equation}
\rho_{A} =  \exp\left( -(\f-\i) K \right).
 \end{equation}
\begin{figure}[hb] 
\begin{center}
\includegraphics[scale=0.5]{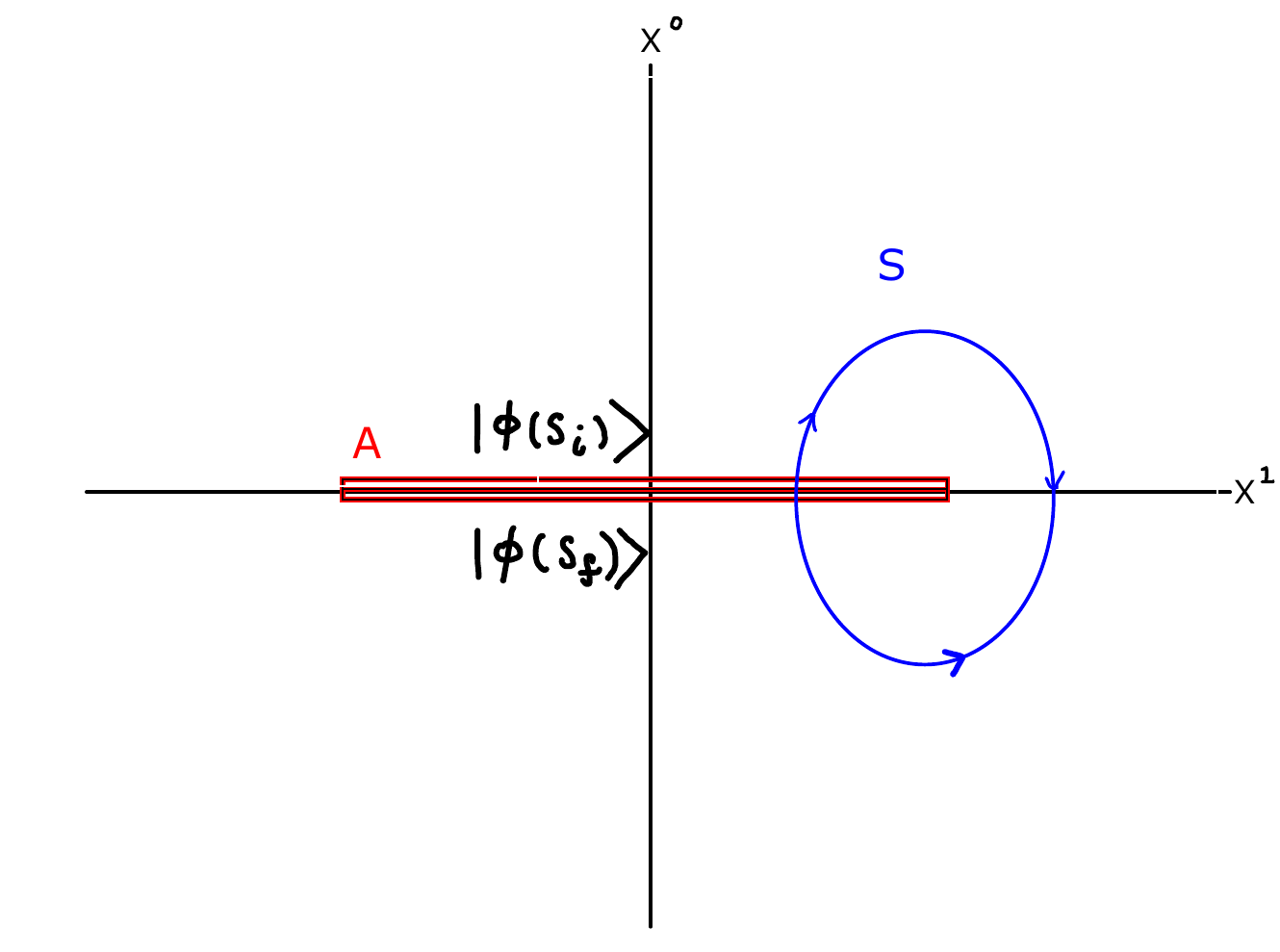}
\caption{{\bf Evaluating \(\rho_{A}\) along Euclidean time s}}
\label{Fig1}
\end{center}
\end{figure}
A well studied situation is the case where the theory is rotationally invariant, and $A ={x^{1}>0 }$ is a half space.   Taking $s$ to be the angular variable on the \(x^{1},x^{0}=t_{E}\) plane, we find the standard result that $K$ is the angular new operator (or the boost generator in Minkowski signature) \cite{unruh1984}. 

From a more general perspective,  $K$  can be viewed as a Killing energy that can be written in terms of the energy momentum tensor.  For any constant $s$ slice \(\Sigma\) we can write
 \begin{align} \label{K}
  K = \int_{\Sigma} T_{ab} k^{a} d \Sigma^{b}, \qquad 
 H_{A} = (\f -\i ) K ,
 \end{align}
where $k^{a} =\frac{dx^{a}}{ds}$ is the Killing vector for the boost and $\{x^{a}\}$ is a set of flat space coordinates.  Choosing to evaluate $K$ on $\Sigma =A$ we find $k^{a} \sim \delta^{a}_{0} $ and $\Sigma^{a} \sim \delta^{a}_{0} $, which reproduces the relation (\ref{Ha}).

Given a spherically symmetric region $A'$ in a Euclidean CFT of any dimension, we will determine the entanglement Hamiltonian for $\rho_{A'}$ by making use of a conformal map $u$ taking $A$ to $A'$, which induces a mapping $\rho_{A} \rightarrow \rho_{A'}=U \rho_{A} U^{-1}$ \footnote{This is essentially a Euclidean version of the arguments in \cite{myers}.}.  The vector field $k'^{a}=\frac{dx^a}{ds'}$ for the new entanglement time $s'$,  is just the image of $k$ under $u$. Thus, the entanglement Hamiltonian for $A'$ is given by (\ref{Ha}) with
\begin{equation}
\label{temp}
  \beta(x) = 2\pi k'^{0}(x),
\end{equation}
where $x\in A$ and the factor of $2\pi$ is simply $s_f-s_i$. We will interpret  $\beta(x)$ as a local ``entanglement'' temperature, that is determined by the shape of $A$ and the background geometry of the CFT. In this interpretation, equation (\ref{Ha}) resembles a density matrix for the original, physical system in local thermal equlibrium with temperature $\beta(x)$. The entanglement entropy is the thermal entropy of this system. It must be emphasized, however, that the appearance of $\beta(x)$ does not correspond to a "real" temperature in the sense that  %\cancel{in any way} 
all inertial observers will find that local observables are at their vacuum values in accordance with \eqref{red}\footnote{However, non-inertial observers whose proper time coincides with $s$ will observe thermal radiation due to the local temperature  \cite{unruh1984}.}. However, the point of view of a local "entanglement temperature" is appealing: indeed $\beta(x)$ must vanish at boundary of the region, signaling a high effective temperature close to the boundary. This behavior may be understood as the statement that the degrees of freedom close to the boundary are the ones most entangled with the external region, and thus have a larger entropy.

Consistent with this interpretation, we have checked that  for two dimensional CFT's in various backgrounds with central charge $c$, the ground state entanglement entropy can be obtained by integrating the equilibrium thermal entropy per unit length 
\begin{equation}
\frac {dS_{thermal}}{dx} = \frac{ c\pi}{3\beta(x)} 
\end{equation}
 over the region $A$ using (\ref{temp}).  Moreover, for excited states \(\beta(x)\) relates the increase in entanglement entropy to an increase in energy inside $A$ via a \emph{local} first  law-like equation:
\begin{equation}\label{density}
d \d S_{A} (x) =  \beta(x) Tr_A(\delta \rho_{A} T_{00} ) dx,
\end{equation}
Here $\frac {d \d S_{A}}{dx} (x)$ is the  local entanglement entropy density\footnote{This is not to be confused with the "entanglement density", introduced in \cite{Nozaki:2013wia} and discussed later in this paper.}  relative to the ground state and $\delta \rho_{A}$ is the variation in the reduced density matrix due to the increase in energy.  To first order in $\d \rho_{A}$, the total increase in entanglement entropy is obtained by  integrating (\ref{density}) over $A$.

Under a \emph{general} variation of the ground state $\rho_{A} \rightarrow \rho_{A} +\delta \rho_{A} $ we find that the  first order change in entanglement entropy is 
\begin{equation}
\label{deltaS}
\delta S_{A} =   Tr_A(\delta \rho_{A} H_{A}).
\end{equation}
 For ground states with other conserved charges \(Q_{a}\) that preserve conformal invariance (e.g. momentum in 1+1 D) , the corresponding charge densities \(q_{a}\) and the associated chemical potentials \(\mu_a\) will appear in the form 
\begin{equation}
H_{A} =\int_{A} \beta(x) (T_{00} - \mu_{a} q_{a}),
\end{equation}
 leading to a generalized first law:
\begin{equation}
d \delta S_{A} (x) =  \beta (x) \d \la T_{00}\ra dx - \beta(x) \mu_{a} \d \la q_{a} \ra dx.
\end{equation}

While preparing this manuscript, a paper \cite{Nozaki:2013vta} was posted where a set of constraint equations for $\d S_{A}$ and an expression for "entanglement density" were derived using AdS/CFT. In section (\ref{Sec:holo}) we provide a CFT derivation of those results in two spacetime dimensions and generalize the constraint equations to arbitrary dimensions\footnote{The constraint equation was recently generalized to holographic CFT's in 3 space-time dimensions in \cite{Bhattacharya:2013fk} }. We will also comment on the relation between our results and those in calculations in \cite{bianchi} and \cite{tak}. 
%Finally, equation \ref{deltaS} is equally valid for variations due to deformations of the shape of $A$ away from spherical symmetry. In this case, the path integral will not be invariant under translations in entanglement time $s$, so $K(s)$ is not conserved and the entanglement Hamiltonian will be non-local. However, given a small perturbation $K_{0} \rightarrow K_{0} + \epsilon K_{1}(s)$ , we can determine the structure of the nonlocal terms by going into an ``interaction picture'' and doing a systematic expansion of equation (\ref{Texp}) in $\epsilon$.
%The paper will be organized as follows. In section 2 we will derive the entanglement Hamiltonian \ref{Ha} and illustrate how to obtain the entanglement temperature in various CFT backgrounds in D=2.  ...%

\section{Path Integral Derivation of the Entanglement Hamiltonian}\label{Sec:Path}
%\subsection{A Proof of (\ref{K}) using the Heat equation}
 Consider a Euclidean QFT on a manifold $M$ and  some spatial region $A$.
The path integral expression for the reduced density matrix on $A$ is similar to the propagator of the theory except that the initial and final states live on the upper and lower edge of a branch cut defined along $A$.  Thus, to switch to a canonical description, it is natural to choose a foliation of $M$ by constant $s$-slices \(\Sigma(s)\) such that the initial/final slice at ``time"$(s_i,s_f)$ lie on the branch cut (see Fig. \ref{Fig1}). The manifold $M$ is then parametrized by coordinates $(s,y^{a})$ where $ y^{a}$ are coordinates on $\Sigma$ .
The reduced density matrix on $A$ in the Schr\"odinger picture is
\begin{eqnarray}\label{rho}
\langle\phi_{0}(s_{f}) | \rho_{A}| \phi_{0}'(s_{i})\rangle = 
 \int D[\phi]e^{-{\cal S}[\phi]}\delta[\phi (s_{f})-\phi_{0}(\f)] \delta[\phi (\i)-\phi_{0}(\i)],
 %\nonumber
\end{eqnarray}
where ${\cal S}[\phi]$ is the action functional. 
To find the entanglement Hamiltonian, we divide the ``time" interval \([\i,\f]\) into small steps \([\N,\n]\) of size \(\Delta s\) and  consider a discretization of the path integral in (\ref{rho}). For notational simplicity we will write $\rho_{A}[\N,\n] = \langle\phi(\N) | \rho_{A}| \phi(\n)\rangle$, so that
\begin{align}\label{dis}
\langle\phi_{0}(s_{f}) | \rho_{A}| \phi_{0}'(s_{i})\rangle = \int d[\phi(s_{N-1})]...  d[\phi(s_{2})] \rho_{A}[\f,s_{N-1}] ...\rho_{A}[\N,\n]... \rho_{A}[s_{2},\i].
\end{align}

Next we will regard the matrix element  $\rho_{A}[\N,s_{n}]$ as a function of the final time $s_{f}$ and final field configuration $\phi(\N, y)$. 
We wish to show that this function satisfies a heat equation
\begin{equation}
\frac{\p}{\p \N} \rn  = - K(\N) \rn
\end{equation}
and identify the operator $K(\N)$. For a given field configuration in the path integral we need to evaluate  $\frac{\p}{\p \N} {\cal S}[ \phi(\N,y), \N]$ at \emph{fixed} $\phi(\N, y)$.  One way of doing this is to keep the final time at $\N$, but transform the background metric by a diffeomorphism that enlarges the proper size of the integration region. 
Explicitly we want a coordinate transformation $s \R s'(s)$  such that  
\begin{align}\label{der}
{\cal S}+ d {\cal S} = \int_{\n}^{\N+ ds }ds \int_{\Sigma(s)} d^{d-1}y \mathcal{L}[g_{ab}, \phi]= \int_{\n}^{\N}ds' \int_{\Sigma(s')} d^{d-1}y \mathcal{L}[g_{ab}+dg_{ab}, \phi],
\end{align}
where $g_{ab}(s, y)$ is the metric on $M $. Therefore,
\begin{align}
d{\cal S}= \int_{\n}^{\N}ds' \int_{\Sigma(s')} d^{d-1}y \frac{\d \mathcal{L}}{\d g_{ab}} dg_{ab}.
\end{align}
In a general coordinate system this transformation and the response of  
 the path integral $\rho_(\N)$  is
\begin{eqnarray} & x^{a} \R x^{a}  = {x^{a}}'-\epsilon^{a}\\ &
d \rho(\N)   = - \frac{1}{2} \int_{[\n,\N]\times \Sigma} \la T_{ab} \ra  \nabla^{(a} \epsilon^{b)} \sqrt{g} d^{d}x = \int_{\Sigma(\N)} \la T_{ab} \ra \epsilon^{b} d\Sigma^{a}.
\end{eqnarray}
Here $\la \ra$ refers to the path integral average on $[\n,\N]$. In the last equality we assumed  the \emph{quantum} conservation law $ \nabla ^{a}\la T_{ab}\ra =0$ and applied the divergence theorem; this means that $T_{ab}$ includes a possible anomalous contribution due to the transformation of the Jacobian in the path integral measure.   
The coordinate transformation that will satisfy equation \ref{der} is 
\begin{equation}
\epsilon^{a} = \frac{dx^{a}}{ds} f(s)  ds,
\end{equation}  
where the function $f(s)$ smoothly goes from 0 to 1 as $s$ goes from $\n$ to $\N$.  This is so that we do not change the lower endpoint of the $s$ integration.

Defining 
\begin{equation}
K(\N)  =  \int_{\Sigma(\N)} \la T_{ab} \ra \frac{dx^{b}}{ds} d\Sigma^{a} ,
\end{equation}
 we find

\begin{align}
\frac{\p}{\p \N} \rho_{A}[\N,\n] = \int D[\phi]e^{-S[\phi]}(-K(\N) )=\langle \phi_{0}(\N) |-K(\N)\rho_{A}| \phi_{0}'(\n)\rangle =-(\hat{K} \rho_{A}) (\N) .
\end{align}
The solution to this heat equation with initial condition $\rho_{A}(\n)=0$ is  \(\rho_{A}[\f,s_{N-1}]= \langle\phi(\N)|1-\Delta s K|\phi(\n)\rangle\).  Inserting this into equation  (\ref{dis}) gives
 \begin{align}
\langle\phi_{0}(s_{f}) | \rho_{A}| \phi_{0}'(s_{i}) \rangle =\int \prod_{n=1}^{N-1} D[\phi(\n)]  \langle\phi(\N)|1-\Delta s K|\phi(\n) \rangle\\
=\langle\phi_{0}(s_{f})|T \exp \left(- \int_{\i}^{\f} K(s) ds \right)| \phi_{0}'(s_{i})\rangle.
\label{K'}
\end{align}

This is the most general form of the entanglement Hamiltonian in a QFT. Since  equation \eqref{K'} only depends on the geometric data provided by the vector field \(\k\) which in turn is determined by the region $A$, it represents a \emph{universal} relation between the entanglement Hamiltonian and the quantum stress tensor.

To recover the local Entanglement Hamitonian (\ref{Ha}), we consider regions $A$ for which $s \rightarrow  s+ d s $ is a spacetime symmetry of the path integral (\ref{rho}) so that $K[s] $ is the corresponding conserved charge.   Since $K$ is independent of $s$, we can evaluate it on any time slice (say at $\i$) and the time ordered product in  (\ref{K'}) reduces to
\begin{equation}\label{local}
\rho_{A} = \exp(-(\f-\i) K(\i)).
\end{equation}
Below we will show that $s \rightarrow  s+ ds $ is indeed a spacetime symmetry of the path integral if $A$ is a half space in a rotationally invariant QFT or a spherical region in a CFT, and we will derive the corresponding local entanglement Hamiltonians.    Here we would like to note that given a small deformation of the region $A$ away from translational or spherical symmetry, one could perform a systematic expansion of equation (\ref{K'}) using the deformed entanglement Hamiltonian \(K_{0} +  \epsilon K_{1} \). To first order in \(\epsilon\) this would just add a perturbation to the local entanglement Hamiltonian which is localized near the boundary of $A$.   A similar strategy can be applied to deformations of the theory away from rotational or conformal invariance.   We leave this for future work. 

\section{Examples of local Entanglement Hamiltonians}
\subsection{Entanglement Hamiltonians in 2D}
\begin{figure} 
\begin{center}
\includegraphics[scale=0.3]{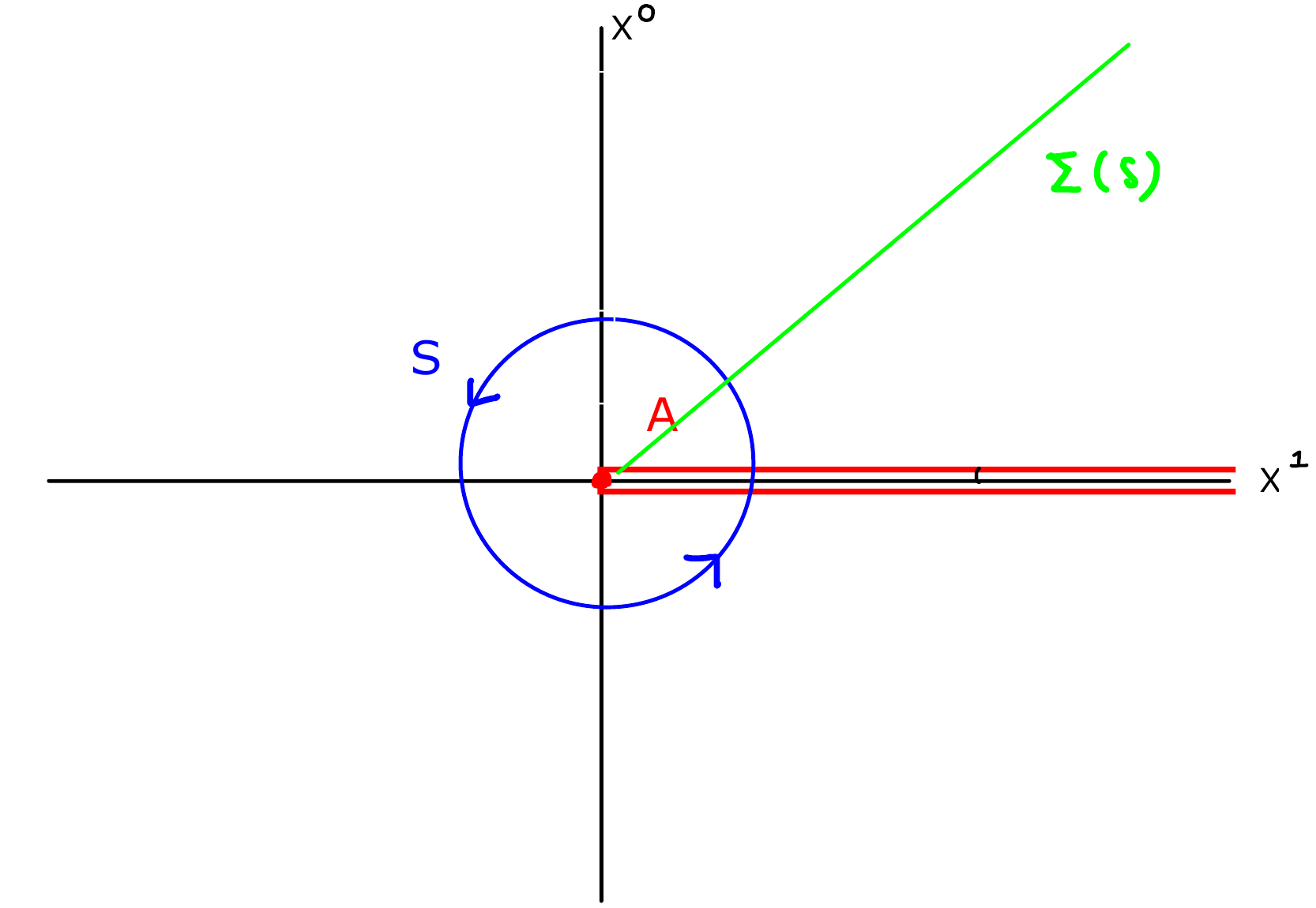}
\caption{{\bf Foliation of the Euclidean plane corresponding to angular quantization}}
\label{Fig2}
\end{center}
\end{figure}

To illustrate how to compute $K$ and its entanglement entropy, we first review the case of a rotationally invariant QFT on $\mathbb{R}^2$ with $A$ the region $A$ being the half line   \(A=\{x^{1}>0\}\) \cite{unruh}.  Since $A$ is mapped to itself by a \(2\pi\) rotation, we choose $s$ to be the angular coordinate on the Euclidean plane so that \(\S(s)\) are rays emanating from the origin as in Figure \ref{Fig2}.

Then 
\begin{equation}
\k\p_a=x^{1}\p_{0}-x^{2}\p_{1}
\end{equation} 
is a Killing vector field generating rotations of the plane.  Since the path integral measure is assumed to be rotationally invariant,  $K$ is just the angular momentum \cite{susskind}
\begin{equation}\label{Rindler}
K= \int_{\S(s=0)} x^{1}T_{00}-x^{0}T_{01}=\int_{A} x^{1}T_{00}.
\end{equation}
The entanglement Hamiltonian is given by equation (\ref{Ha}) with the entanglement temperature
\begin{equation}
\beta= 2\pi x^1.
\end{equation}
    Upon Wick rotating \(s \R i s\), the circular flow generated by $K$ becomes hyperbolas representing the worldlines of uniformly accelerated observers, and \(\beta(x)\) is the proper temperature they experience.  Thus in Minkowski signature $K$ is the boost generator. The form of the entanglement Hamiltonian implies that \(\rho_{A}\) represents an ensemble with the physical energy density \(T_{00}\) in local thermal equilibrium with local temperature \(\beta(x)\); its entanglement entropy is therefore just the thermal entropy, obtained by integrating the thermal entropy density \(\frac {dS_{thermal}}{dx}\) over $A$ \cite{susskind}. 

   In particular, for a CFT with central charge $c$,  it is  well known that \cite{hlw} 
\begin{equation}
\frac {dS_{thermal}}{dx}=\frac{c \pi}{3\beta(x)}
\end{equation}
so the entanglement entropy is
\begin{equation}
S_{A}= \int_{\delta}^{L}dx \frac{c \pi}{6 x}=\frac{c}{6} Log \frac{L}{\delta},
\end{equation}
where we have introduced a UV and IR cutoff on $A$ restricting the integration to \([\d,L]\).  The local temperature is higher near the boundary of $A$ and diverges at $x=0$ due to the zero of the vector field, which is also the singularity of the foliation defined by $s$.   As a result, most of the contribution to the entanglement entropy arises from near the edge. % Is it possible to treat the edge as a source of charge for the vector field?  In the context of the replica trick, the edge is the location of the twists operators, which can be interpreted as putting a dirac string for a gauge field along the edge... %

For a CFT on $\mathbb{R}^2$ we can easily generalize the previous results to an arbitrary interval $A'=[u,v]$.  Let \(z=x^{1}+i x^{0}\) so that \(\frac{dz}{ds}=i z\) is the rotational vector field appropriate to the region $A$ discussed previously.  The  conformal map
\begin{equation} \label{cm}
 z=-\frac{w-u}{w-v}
 \end{equation}
 induces a transformation $U$ on the reduced density matrices:
\begin{equation}\label{U}
\rho_{A} \R \rho_{A'}= U \rho_{A} U^{-1},
\end{equation}
by transforming the boundary conditions of the path integral. The path integral measure is conformally invariant because there is no anomaly in flat space. Meanwhile, the vector field \(\frac{dz}{ds'}\) is mapped to
\begin{equation}
\frac{dw}{ds'}  =  \frac{dw}{dz} \frac{dz}{ds'} =\frac{i(w-u)(w-v)}{u-v}.
\end{equation}
It is clear that the periodic flow defined by this vector field will evolve $A' \to  A'$.   Moreover, the transformation \(w \R w + \frac{dw}{ds'} d s'\) is a symmetry of the CFT on the $w$ plane, because it can be decomposed into a combination of a conformal transformation between $z$ and $w$, and an ordinary rotation on the $z$ plane.

\begin{figure}
\begin{center}
\includegraphics[scale=0.8]{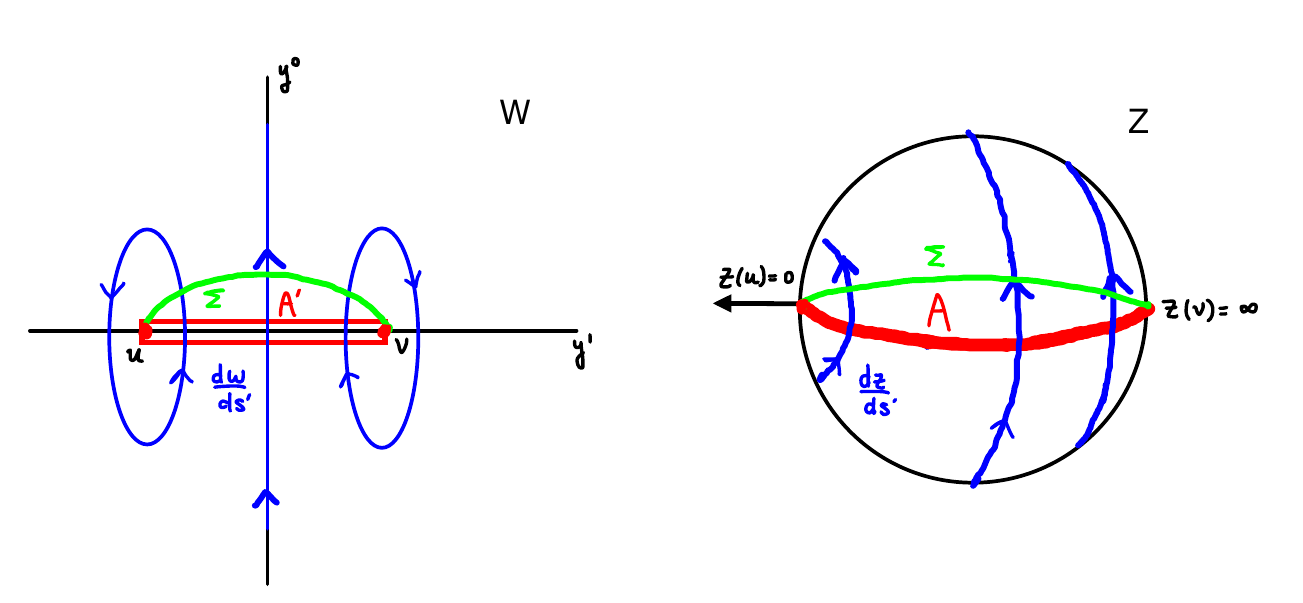}
\caption{{\bf A rotation on the z plane (represented as a Riemann sphere) is mapped to a conformal rotation on the w plane}}
\label{default}
\end{center}
\end{figure}

Thus, the entanglement Hamiltonian for \(\rho_{A'}\) is
\begin{equation} \label{HaR2}
H_{A'} =  \int_{A} 2\pi \frac{(y^{1}-u)(y^{1}-v)}{u-v}T_{00} dy ,
\end{equation}
where we defined \(w= y^{1}+iy^{0}\) and evaluated the integral along $A$ for convenience.
As before, the entanglement entropy is obtained from integrating \(\frac {dS_{thermal}}{dx}\) using the entanglement temperature 
\begin{equation}
\beta(y) =  2\pi \frac{(y^{1}-u)(y^{1}-v)}{u-v}.
\end{equation}
 This gives  
\begin{equation}
S_{A'} = \frac{c}{3} log \frac{v-u}{\d},
 \end{equation}
as expected\footnote{ Note that even though \(Tr_A(\rho_{A} log \rho_{A})\) is invariant under the similarity transformation (\ref{U}) of the reduced density matrix,  we get a different result for the entanglement entropy of  \(\rho_{A'}\) because we have to transform the regularized boundary of $A$.  }.

For a CFT at finite temperature (\(w \sim w+i \beta'\)) or on a spatial circle (\(w \sim w+ L\)), we can similarly derive the entanglement Hamiltonian by finding the conformal map from the $z$-plane to  \(\mathbb{R} \times S^{1}\) or \(S^{1} \times \mathbb{R}\).  Given $A' = [-l,l] \times \{0\}$, the conformal map and entanglement temperature for a CFT at the (ordinary) temperature $\beta'$ is
\begin{align}
\label{ft}
z= \frac{-\exp\bigg(\frac{2\pi w}{\beta'}\bigg) + \exp\bigg(-\frac{2\pi l}{\beta'}\bigg) }{\exp\bigg(\frac{2\pi w}{\beta'}\bigg)- \exp\bigg(\frac{2\pi l}{\beta'}\bigg)}, \qquad 
\beta =  2 \beta' {\rm csch}\left(\frac{ 2l \pi}{\beta'}\right) \sinh\left(\frac{\pi (l - y)}{\beta'}\right) \sinh\left(\frac{\pi (l+y}{\beta'}\right).
\end{align}
The results for a CFT at finite size can be obtained from equation (\ref{ft}) by the substitution \( \beta' \R i L\).
Below we summarize the results for the entanglement temperature and entanglement entropy obtained by integrating the thermal entropy density in various CFT backgrounds.  
% Simpler to consider A=[-l,l] for all cases %
\begin{table}[htdp]
\caption{Entanglement Temperature for $A=[-l,l]$ in different CFT backgrounds}
\begin{center}
\begin{tabular}{|l |c| c | r }
\hline
CFT Background & Entanglement Temperature \(\beta(y)\) & Entanglement Entropy \(S_{A}=\int_{A} \frac{\pi c}{3 \beta (y)}\)\\ \hline
Zero Temp. and infinite size: $M=\mathbb{R}^{2}$&\(\frac{2\pi (l^{2}-y^{2})}{2l}\)&\(\frac{c}{3}\ln \frac{l}{\d}\) \\ \hline
Finite temperature $\beta': M= \mathbb{R} \times S^{1}$& $ 2 \beta' {\rm csch}\left(\frac{  2l \pi}{\beta'}\right)\sinh\left(\frac{\pi (l -
     y)}{\beta'}\right) \sinh\left(\frac{\pi (l+ y)}{\beta'}\right) $ & $\frac{c}{3} \ln\left( \frac{\beta'}{\pi \d} \sinh(\frac{2\pi l}{\beta'}) \right)$\\ \hline
Finite Size $L: M=S^{1} \times \mathbb{R}$&\(2 L {\rm csc}\left(\frac{2l \pi}{L}\right) \sin\left(\frac{\pi (l - y)}{L}\right) \sin\left(\frac{\pi (l+y)}{L}\right) $&\(\frac{c}{3} \ln\left(( \frac{L}{\pi \d} \sin(\frac{2\pi l}{L}) \right)\)\\
\hline
 
%Finite Size $L: M=T^{2}(\tau)$ & $ 2 \beta' {\rm csch}\left(\frac{  2l \pi}{\beta'}\right)\sinh\left(\frac{\pi (2l -
     %y)}{\beta'}\right) \sinh\left(\frac{\pi y}{\beta'}\right) $ &\(\frac{c}{3} \ln\left(( \frac{L}{\pi \d} \sin(\frac{2\pi l}{L}) \right)\)\\

\end{tabular}
\end{center}
\label{default}
\end{table}%

The results for the entanglement entropies were derived previously using the replica trick, \cite{calabrese2004entanglement, calabrese2009entanglement, hlw,wl}, and serve as a check on our results for the entanglement temperature and  Hamiltonian. 

%%%%%%%%%%%%%%%%%%%%%%%%%%%%%%%%%%%%%%%%%%%%%%%%%%%%%%%%%%%%%%%%%%%%%%%%%%%%%5
\subsection{Entanglement Hamiltonians in higher dimensions}
Here we generalize the results of the previous section to spherical entangling surfaces in dimensions \(d>2\).   As before, we first consider a rotationally invariant CFT on $\mathbb{R}^d$ with \(A=\{x^{1} > 0 \} \).  We choose polar coordinates on the \( x^{1},x^{0}\) plane \( x^{1}= z \cos (\frac{s}{l})\), \(x^{0} =z \sin (\frac{s}{l})\), so the flat metric is
\begin{align}
d\tau^{2} = (\frac{z}{l})^{2}ds^{2} +dz^{2} + d\vec{x}^2  .
\end{align}
At this point $l$ is an arbitrary length parameter introduced to make $s$ dimensionful.
Then the result (\ref{Rindler}) for the entanglement Hamiltonian of \(\rho_{A}\) is still valid. Now we map $\mathbb{R}^{d} \R H^{d-1} \times S^{1}$, by multiplying the metric above by a conformal factor  \( (\frac{l}{z})^2\).
\begin{equation}
d\tau^{2}_{H^{d-1}\times S^{1}}=ds^{2} + (\frac{l}{z})^2 (dz^{2} + d\vec{x}^2  ).
\end{equation}
The \(H^{d-1}\) factor refers to hyperbolic space, which is the image of the half space $A$. Thus we see that \(\rho_{A}\) is transformed into a thermal density matrix \(\rho_{H^{d-1}}\) on hyperbolic space.  Since this conformal map does not change the original coordinates on $\mathbb{R}^{d}$, the vector field generated by the new entanglement Hamiltonian is just \(\frac{\p}{\p s} \).

Now consider a new reduced density matrix \(\rho_{A'}\) for a ball of radius $l$.   We will obtain the entanglement Hamiltonian \(H_{A'}\) by mapping \(\rho_{H^{d-1}} \R \rho_{A'} \) as follows.  First we choose coordinates \( ( u,\Omega_{d-2},s) \) on  \(H^{d-1} \times S^{1}\) and spherical coordinates \((r, \Omega_{d-2},t)\) on $\mathbb{R}^{d}$  so that the metrics are
\begin{align}
d\tau^{2}_{H^{d-1} \times S^{1}} = ds^{2} + R^{2}(du^{2}+\sinh(u)^{2} d \Omega^{2}_{d-2}),\\
d\tau^{2}_{R^{d}} = dt^{2}+ dr^{2} +r^{2} d\Omega^{2}_{d-2}.
\end{align}
Then, defining complex coordinates \(\sigma=u+i \frac{s}{l}\) and \(w=r+it\) on the respective two dimensional slices, we consider the mapping introduced in  \cite{hung}
\begin{equation}\label{hung}
e^{-\sigma}= \frac{l-w}{l+w}.
\end{equation}
This is an analogue of equation  (\ref{cm}) mapping \(\rho_{A'}\R \rho_{H^{d-1}}\). The entanglement vector field and entanglement Hamiltonian is
\begin{align}
\frac{dw}{ds}=\frac{dw}{d\sigma}\frac{d\sigma}{ds}= i \frac{l^{2}-r^{2}}{2l}, \qquad 
H_{A}= 2 \pi \int_{A} \frac{l^{2}-r^{2}}{2l} T_{00}.
\end{align}
This agrees with the result of \cite{myers}, where a Minkowski signature version of the conformal mapping (\ref{hung}) was used to derive the entanglement Hamiltonian.

%%%%%%%%%% %%%%%%%%%%%%%%%%%%%%%%%%%%%%%%%%%%%%%%%%%%%%%%%%%%%%%%%%%%%%%%%%%%%%%%%% %%%%%%%%%%%%%%%%%%%%%%%%%%%%%%%%%%%%%%%%%%%%%%%%%%%%%%%%%%%%%%%%%%%%%%%%
\section{CFT derivation of Entanglement Entropy for excited states}
Consider a state $|\psi \ra$ in a QFT  in $\mathbb{R}^{1,d-1}$ with a density matrix \(\rho^{0}=|\psi \rangle\langle \psi|\).  As in  \cite{bianchi} we make a small perturbation  $\rho= \rho^{0}+\delta \rho$ and consider the entanglement entropy of a region $A$. Expanding to first order in \(\delta \rho_{A} \) we find
\begin{equation}
S_{A}=-Tr_{A}(\rho_{A} \ln \rho_{A})=-Tr_{A} (\rho^{0}_{A}\ln \rho^{0}_{A})-Tr_{A}(\delta\rho^{0}_{A} \ln \rho^{0}_{A})-Tr_{A}(\delta \rho_{A}),
\end{equation}
where $\delta\rho_A=Tr_B(\delta\rho)$.
The normalization \(Tr(\rho_{A})=Tr_A(\rho^{0}_{A})=1\) implies  \(Tr(\delta \rho_{A})=0\), so the first order change in entanglement entropy due to the perturbation $\d \rho $ is simply
\begin{equation} \label{EE}
\d S_{A}=-Tr_{A}(\delta\rho_{A} \ln \rho^{0}_{A})=Tr_{A}(\delta \rho_{A} H_{A}).
\end{equation} 
Note that there is also a term proportional to ${\rm Tr}(\delta \rho)$ which vanishes due to the normalization \(Tr(\rho)=1\). When the state $\rho^{0}$ is the ground state, we will refer to 
\(\d S_{A}\) as the \emph{renormalized} entanglement entropy\footnote{This is only a first order approximation to the renormalized entropy, but we will just call it renormalized entropy for short.} \cite{hlw}. It is just the increase in ``entanglement energy'' of the new state, measured according to the ground state entanglement Hamiltonian.    However we emphasize that equation (\ref{EE}) applies to an \emph{arbitrary} deformation \(\d \rho\) for  \emph{any} initial state $\rho^{0}$.
When the region $A$ is a half space in a QFT or a spherical ball in a CFT, we can use the entanglement temperatures previously derived to obtain \(H_{A}\) for the ground state as in equation  (\ref{Ha}).  From equation (\ref{EE}) we have:
\begin{align} \label{EE1}
\d S_{A}=Tr_{A}(\delta \rho_{A} \int_{A}\beta(x)T_{00}(x))=\int_{A}\beta(x)Tr(\delta \rho T_{00}(x)):=\int_{A}\beta(x) \d \la T_{00}(x) \ra
\end{align}
In the second to last equality, we noted that the operator \(T_{00}(x)\) is only being evaluated inside $A$ so that \(\delta \rho_{A} \) can be replaced with \(\delta \rho \). Note that in (\ref{EE}) the operator $\d \rho_{A}$ and $H_{A}$ are defined on a subregion $A$ with boundaries, which implies boundary conditions have to be imposed at $\p A$ on their quantization.  On the other hand, in (\ref{EE1}) the operator $T_{00}$ is interpreted as the energy density quantized with the boundary conditions appropriate to the \emph{whole} space; we have merely chosen to \emph{evaluate} it inside $A$. These two interpretation must agree by the definition of the reduced density matrix.  As a check, in appendix B we will show that for a particular excitation of a free scalar field with non-uniform energy density,   (\ref{EE1})  and (\ref{EE}) do indeed give the same result for  \(\d S_{A}\).

When \(\d \la T_{00} \ra \) is spatially uniform\footnote{Since our entanglement Hamiltonian was derived for a CFT on $\mathbb{R}^d$, we will assume the energy density starts to die off somewhere outside $A$, in order for the energy to be finite.} inside $A$, we can remove it from the integration, so that
\begin{equation} \label{EE2}
\d S_{A}= \beta_{0}  \d \la T_{00} \ra
Vol(A) := \beta_{0} \d E_{A},
\end{equation}
where $\d E_{A} = \d \la T_{00} \ra {\rm Vol}(A)$  is the excitation energy inside region $A$, and \(\beta_{0} \) is the average entanglement temperature inside $A$
\begin{equation} \label{beta0}
\beta_0=\frac{\int_A\beta(x)}{{\rm Vol}(A)}.
\end{equation}
 When the region $A$ has radius $l$, we find\footnote{ As already noted in \cite{tak}, this is also consistent with the computation of \(\d S_{A} \) for primary states of a two dimensional CFT which was performed in \cite{Berganza:2011mh} via the replica trick.} \(\beta_{0} = \frac{2\pi}{d+1} l\) in agreement with the result of \cite{tak}. However, we note that the holographic results of \cite{tak} only strictly apply to nonabelian gauge theories with holographic duals, at large N and assuming a small region $A$ (i.e. for small radius $l$), whereas our result is valid to order $O(\delta \rho)$ for any CFT and any radius $l$. We also note that there is a discrepancy between our results when \(\d \la T_{00}\ra \) is spatially varying.  Given a state with  \( \d \la T_{00} \ra = \sum_{n=0}^{\infty} a_{n} r^{n}\)  in a $d>2$ dimensional CFT\footnote {We will explain the restriction to $d>2$ in the section VI.},  we find 

\begin{equation}\label{Entropy of excited sphere}
\d S_{A}= 2 \pi  {\rm Vol}(S^{d-2}) \sum_{n=0} \frac{a_{n}l^{d+n}}{(d+n)^{2} -1} 
\end{equation}
which disagrees with the holographic calculation of the same quantity  in equation (20) of \cite{tak}.   In section IV, we will discuss the holographic version of eq \ref{EE} and speculate on a possible source of the discrepancy.   As noted earlier, we have checked in appendix B that our results (\ref{EE}) and (\ref{EE1}) are consistent for a \emph{non-uniform} excitation of a free scalar field, where $\d S_{A}$ can be computed explicitly.  

 \section{A generalized first law for entanglement entropy}
%%%%%%%%%%%%%%%%%%%%%%%%%%%%%%%%%%%%%%%%%%%%%%%%%%%%%%%%%%%%%%%%%%%%%%%%%%%%
Equation (\ref{EE1}) resembles a local first law of thermodynamics inside the region $A$:
 \begin{equation}
 \label{flaw}
d \delta S_{A}(x)=\beta(x) \d  \la T_{00}(x) \ra dx.
\end{equation}
When other conserved charges are present,  a generalization of equation (\ref{flaw}) can be derived as follows.
Consider a state at finite temperature $T$ and with conserved charges \(Q_{a}\)  that preserve conformal invariance and chemical potentials \(\mu_{a}\) weighted with the following density matrix
 \begin{equation}
 \rho =\frac{\exp\bigg(-\frac{(H- \mu_{a}Q_{a})}{T}\bigg)}{Z}.
 \end{equation}

After tracing over the complement of $A$ we arrive at a  path integral representation of \(\rho_{A}\) similar to the one given in equation (\ref{rho}), except that
adding the charges has effectively shifted our Hamiltonian from $H$ to \(H' =H- \mu_{a}Q_{a}\).
The corresponding shift in the energy density is \(T_{00}' = T_{00}-\mu_{a}q_{a} \), where we introduced the charge densities $q_{a}$ by $Q_a:= \int_{space} q_{a}d^{d-1}x$.   Going through the same path integral derivation as in section \ref{Sec:Path}, we would reproduce equation (\ref{Ha}) with \(T_{00}\) replaced by \(T_{00}'\). Under a deformation \( \delta \rho \) that changes the charge densities and energy inside $A$,  equation  (\ref{flaw}) now becomes
 \begin{equation}  \label{gflaw}
d \delta S_{A}(x)=\beta(x) \delta  \la T'_{00}(x) \ra  dx =\beta(x) \{\d \la T_{00}(x) \ra dx  - \mu_{a} \d \la q_{a}(x) \ra dx\}
 \end{equation}

A simple way to check the above argument  for the entanglement Hamiltonian  leading to equation  (\ref {gflaw}) is to consider a state \(\rho \sim \exp[- \beta'(H-\mu P)]\) for a two dimensional CFT with total central charge $c$.   In this case the conserved Virasoro charges are the Hamiltonian \(H=L_{0} +\widebar{L_{0}} -\frac{c}{12} \) and momentum \(P=L_{0} -\widebar{L_{0}} \).   The entanglement Hamiltonian for an interval $A=[0,l]$ is
\begin{equation} \label{Ha'}
H_{A}= \int_{0}^{l} \beta(x)(T_{00}-\mu T_{01}) dx = \int_{0}^{l} \beta(x)(1-\mu)T_{++} + \beta(x)(1+\mu)T_{--},
\end{equation}
where \(T_{\pm \pm} =\frac{1}{2}(T_{00} \pm T_{01})\) are the right and left moving components of the stress tensor, and \(\beta(x)\) is the entanglement temperature (\ref{ft}) for a CFT at finite temperature\footnote{Technically, to get a discrete spectrum for $P$ we should put the CFT on a spatial \(S^{1}\) of length $L$. Here we will assume \(\beta' >> L\), so that we can ignore the periodicity along $L$ in computing the entanglement temperature.} \(\beta'\). The operator in equation (\ref{Ha'}) is the sum of two \emph{commuting} entanglement Hamiltonians corresponding to  non-interacting  ensembles at finite (ordinary) temperature \(\beta_{\pm} =\beta'(1\pm \mu)\) and with energy density \(T_{\pm \pm }\).   Assuming that the left and right central charges are equal, each ensemble has an effective central charge of \(\frac{c}{2}\) . Thus the entanglement entropy is:
\begin{align}\label{EE'}
S_{A}= \frac{c}{6} \ln\left( \frac{\beta_{+}}{\pi \d} \sinh(\frac{\pi l}{\beta_{+}}) \right) + \frac{c}{6} \ln\left( \frac{\beta_{-}}{\pi \d} \sinh(\frac{\pi l}{\beta_{-}}) \right).
\end{align}
This agrees with the result of \cite{hubeny2007covariant} obtained via the replica trick and holographic calculations.    %Moreover,eq \ref{EE'} naturally generalizes to the case when the left and right moving sector has different central charges ... just be replacing c/6 by c

%%%%%%%%%%%%%%%%%%%%%%%%%%%%%%%%%%%%%%%%
\section{Holographic derivation and discussion of related papers}\label{Sec:holo}

According to the holographic prescription of \cite{rt}, the entanglement entropy for  a state \(|\psi \rangle\) in a region $A$ of a $d$-dimensional CFT with a holographic dual gravity theory is 
\begin{equation} \label{RT}
S_{A}=\frac{Area(\gamma_{A})}{4 G},
\end{equation}
where \(\gamma\) is a minimal surface, anchored on \(\p A\), in the bulk spacetime representing the gravity dual of the corresponding CFT, $G$ is the bulk Newton's constant.  The geometry dual to the ground state  in the CFT corresponds to pure AdS
\begin{equation}
d\tau^{2} = (\frac{R}{z})^{2}(-dt^{2} + dz^{2} + r^{2}d\Omega^{2}_{d-2}),
\end{equation}
and the minimal surface for  \(A=\{r=l\}\) is a half sphere extending into the bulk: \(\gamma_{A} = \{ r^{2}=l^{2} -z^{2}\} \).

For general excited states, it is difficult to find the exact bulk metric and compute the minimal surface. However, just as in the CFT computation of the previous section, a drastic simplification occurs if we consider only the first order deformation of the entanglement entropy, which is proportional to  the variation of area functional :

\begin{equation} \label{dA}
\d Area(\gamma_{A})= \d \int_{\gamma_{A}} \sqrt{g} =\int_{\gamma_{A}}\d \sqrt{g}.
\end{equation}
In the last equality, we observed that the  area variation due to the deformation of the surface \(\gamma_{A}\) vanishes by the definition of a minimal surface.  Thus, the area variation is entirely due to the change in the metric, and there is no need to solve for the minimal surface in the new geometry.   Comparing this equation to (\ref{EE2}), we see that \(\d \rho_{A} \) corresponds to the deformation of the metric while \(H_{A}\) corresponds to the ground state minimal surface.  The second fact is less obvious from the usual AdS/CFT correspondence, but it is consistent with ideas proposed in  \cite{myers}. In reference \cite{myers} it was shown that for spherical regions A, there exists a foliation of AdS by hyperbolic slices \(\mathcal{H}=H^{d-1}\times R \)  such that one of the slices is a  causal horizon \(\gamma_{A}' \)that is anchored on \(\p A\).  Since a horizon is also a minimal surface, we can identify \(\gamma_{A}'=\gamma_{A}\).  The new foliation of AdS is dual to a CFT on the boundary slice \(\mathcal{H} \), which is in a thermal state that is conformally related to \(\rho_{A}\) . It is thus tempting to identify the foliation of AdS  and the associated horizon \( \gamma_{A} \) with the reduced density matrix \( \rho_{A} \) and therefore \(H_{A}\).

As in \cite{tak}\footnote{see also \cite{allahbakhsi2013e} for an extension of results in \cite{tak}} we consider an excited state with energy density   \footnote{To facilitate comparisons with \cite{tak}, in this section  we write $ \d \la T_{00} \ra = \la T_{00} \ra $,  with the understanding that the energy density in the latter expression is normal ordered so as to subtract the vacuum energy.  Note that there is a typo in eq. (2) of \cite{tak} where $d$ was replaced with $d-1$.} \( \langle T_{00} \rangle = \frac{d R^{d-1} m }{16 \pi G} \). As established  in ref. \cite{sk}, the holographic stress tensor associated with this energy density and the boundary metric determines  the \emph{asymptotic} form of the bulk metric near the boundary at \(z\sim 0\) to be:

\begin{align}\label{FG}
d\tau^{2} = (\frac{R}{z})^{2}(- g^{-1}(z) dt^{2} + g(z)dz^{2} + r^{2}d\Omega^{2}_{d-2}), \quad {\rm with} \quad 
g(z)=1+mz^{d}+ ...
\end{align}
where the ellipsis denotes higher order terms in $z$.   In this approximation, the first order variation  of the entanglement entropy for spherical regions A is
\begin{eqnarray}
\frac{\d S'_{A}}{\d m}\bigg|_{m=0} \d m &=&\frac{\d Area(\gamma_{A})}{4G} \bigg|_{m=0}= R^{d-1}\Omega_{d-2} \int_{0}^{l} \frac{r(z)^{d-2}}{z^{d-1}} \d \sqrt{g(z) + r'(z)^{2}} \nonumber \\
&=& \beta_{0} \d E_{A},
\end{eqnarray}
where we evaluated the integral along the half sphere \( r^{2}=l^{2} -z^{2}\) corresponding to the ground state at $m=0$,  $\beta_{0}=\frac{2\pi}{d+1} l$, and $\d E_{A}$ is defined as in the section IV.  The notation $\d S'_{A}$ is a reminder of the additional approximation due to the expansion \eqref{FG}, where  sub-leading in terms in $z$ were dropped.  However, in this case, this approximation (truncation) leads to a result which agrees with the field theoretic one in eq. \eqref{beta0}

Next, we consider a  a non-uniform state with energy density $\la T_{00}\ra =  \frac{d R^{d-1} m }{16 \pi G} \sum_{n\geq 0} c_n r^n $ in a $d>2$ dimensional CFT.
Note that this state is not allowed $d=2$ spacetime dimensions, because the energy density has to satisfy a wave equation, as explained  later in this section.    The dual metric has the same form as in \eqref{FG} with  
\begin{align}
g(z)=1+m z^d \sum_{n\geq 0} c_n r^n+ \dots,
\end{align}

using \eqref{dA} we find:
\begin{eqnarray}
\delta S'_{A}= {m l^d  R^{d-1}{\rm Vol}(S^{d-2})\over 8G}\sum_{n\geq 0} {c_n l^n \over 1+d+n}.
\end{eqnarray}
The above expression reproduces and generalizes the results in \cite{tak}, without recourse to a an explicit evaluation of the minimal surfaces. This time, we note that above $\delta S'_{A}$ differs from our result \eqref{Entropy of excited sphere} for the entropy of a sphere, although both are supposed to represent entropy of a system with the same non-uniform energy density. 

In  \cite{tak},  use of equation (\ref{FG}) was justified by taking the small region limit, that is, the  \( l\R 0 \) limit in which  \(\gamma_{A} \) approaches the \(z=0\) boundary.   However, neglect of higher order terms in $z$, while not affecting the energy density \(\langle T_{00} \rangle \), may affect the computed entropy. For example, adding a correction of the form   $m z^{d+k}r^\mu$ will yields, using \eqref{dA}, a contribution proportional to  $l^{d+k+\mu}$ to the holographic entropy. Neglect of such terms 
%Taking this limit in our CFT calculation in (\ref{EE1}) corresponds to choosing treating  \(\langle T_{00} \rangle \) as constant over the infinitesmal integration region A. 
may be the reason that our results agree with those of \cite{tak} only for the case of uniform energy density. In this way, our result provides an easy consistency check for the $z\rightarrow 0$ limit metric used in holographic calculations. 
\subsection{Dynamical  equations  for  entanglement entropy and entanglement density}

While this project was being completed, we noticed a recent paper \cite{Nozaki:2013vta} where a set of dynamical equations were derived for \(\d S_{A}\) in the case of  time dependent excited states by using the holographic formula (\ref{RT}).  In $d=2$ spacetime dimensions they are:
\begin{align} \label{D'}
(\p_{t}^{2} - \p_{\xi}^{2}) \d S_{A} (\xi, l, t) =0 \\
 (\frac{\p_{l}^{2}}{4} - \frac{\p_{t}^{2}}{4} -\frac{1}{2l^{2}} ) \d S_{A} (\xi,l,t)=0 
 \label{D''}
\end{align}
where \(A= [ \xi- l, \xi +l ]\). In the holographic setting these equations arose from solving Einstein's equations perturbatively to determine the evolution of the metric for the excited state. Here we will provide a simple field theoretic derivation of these equations. First note that in terms of the variable \(x'=x-\xi\), the renormalized entanglement entropy  for  a CFT on a plane is
\begin{equation}\label{REE}
 \d S_{A} = 2\pi\int_{-l}^{l} dx' \frac{l^{2}-x'^{2}}{2l^{2}} \langle T_{00} \rangle(x'+\xi,t),
\end{equation}
so the entanglement temperature is independent of \(\xi\).  Thus,
\begin{equation}
(\p_{t}^{2} - \p_{\xi}^{2}) \d S_{A} (\xi, l, t) =2\pi\int_{-l}^{l} dx' \frac{l^{2}-x'^{2}}{2l^{2}} (\p_{t}^{2} - \p_{\xi}^{2})\langle T_{00} \rangle(x'+\xi,t)=0,
\end{equation}
where in the last equality we used the fact that in $d=2$ the conservation of the energy momentum tensor combined with its tracelessness imply that \(T_{00}= T_{++} + T_{--}\) is a sum of left and right movers, and therefore satisfy the wave equation.  The second equation (\ref{D'})can be obtained straightforwardly by applying the differential operator to (\ref{REE}) and integrating by parts using \(\p^{2}_{t}T_{00}=-\p_{\xi}^{2}T_{00}=-\p_{x'}^{2}T_{00}\).
As in \cite{Nozaki:2013vta}, we can also generalize and (\ref{D'}) to the case when we couple an operator \(O(x,t) \) to a  source \(J(x,t)\)  so that our physical Hamiltonian is deformed to \(H' = H - \int J O d^{d-1}x\).  Provided that $O(x,t) $ preserves conformal symmetry, this deformation changes the ground state Hamiltonian by deforming the energy density \(T_{00}\R T_{00}'= T_{00}- J O\) in \ref{Ha}. The equations (\ref{D'}) are now modified by source terms that arise form the differential operators hitting \(J(x,t)O(x,t)\).
Thus
\begin{align}\label{source}
(\p_{t}^{2} - \p_{\xi}^{2}) \d S_{A} (\xi, l, t) = \int_{-l}^{l} \beta(x',l)(\p_{t}^{2} - \p_{\xi}^{2} ) (J(x'+\xi,t) \la O(x'+\xi,t)\ra_{J}), \\
(\frac{\p_{l}^{2}}{4} - \frac{\p_{t}^{2}}{4} -\frac{1}{2l^{2}} ) \d S_{A} (\xi,l,t)= - \int_{-l}^{l}\beta(x',l) \frac{\p_{t}^{2}}{4}(J(x'+\xi,t)\la O(x'+\xi,t) \ra_{J} ),
\label{source'}
\end{align}
with
\begin{equation}
\beta(x', l)=2\pi\frac{l^2-{x'}^2}{2l^2}.
\end{equation}
To facilitate a comparison with the result of \cite{Nozaki:2013vta}, we  take the Fourier transform of  \( \la O(x'+\xi,t) \ra_{J}\)  and make explicit the dependence of  \(J(x'+\xi,t) \) on \( \la O(k_{1},w_{1})\ra_{J}\):
\begin{align}
 \la O(x,t)\ra_{J}  =  \int d \omega_{1} \int dk_{1}   \la O(k_{1},\omega_{1})\ra_{J} e^{i(k_{1}\xi+\omega_{1}t)} e^{ik_{1}x'}, \\
 J(x'+\xi,t) = \int d \omega_{2} \int dk_{2}  f(k_{2},\omega_{2})   \la O(k_{2},\omega_{2})\ra_{J} e^{i(k_{2}\xi+\omega_{2}t)} e^{ik_{2}x'}.
 \end{align}
 Above we chose the source $J$ corresponding to the perturbation of the bulk scalar given in  equation (3.17) of  \cite{Nozaki:2013vta}. Inserting these in
 (\ref{source})  and integrating over $x'$ gives equations of the form
 \begin{align}
 (\p_{t}^{2} - \p_{\xi}^{2}) \d S_{A} (\xi, l, t) =\int d\omega_{1} \int d\omega_{2} \int dk_{1} \int dk_{2}  F(k_{1},k_{2},\omega_{1},\omega_{2},l ) \la O(k_{1},\omega_{2})\ra_{J}  \la O(k_{2},\omega_{2})\ra_{J} e^{i((k_{1}+k_{2})\xi+(\omega_{1}+\omega_{2})t)},
\end{align}
 and similarly (\ref{source'}).  These equations have the same form as (3.22) and (3.23) of \cite{Nozaki:2013vta}, which were interpreted as the holographic dual to the perturbative Einstein's equations with the right hand side serving as the matter source.

In general dimensions,we can derive a constraint equation similar to \ref{D''} for a ball $A$ of radius $l$ centered on $\vec{\xi}$ :
\begin{align} \label {cst}
(\p_{l}^{2} -  (d-2) \frac{\p_{l}}{l} - \nabla^{2}_{\xi} -\frac{d}{l^{3}} ) \d S_{A} (\vec{\xi},l,t)=0 
\end{align} 
As in the case of 2 dimensions , this can be verified straightforwardly by applying the  differential operator above to the expression for  $\d S_{A}$ in \eqref{REE} and integrating by parts after noting that:
\begin{equation}\label{green}
\int_{A} \beta (r) \nabla^{2}_{\vec{\xi}}T_{00} (\vec{\xi}+\vec{r}) dr d\Omega=\int_{A} \beta (r) \nabla^{2}_{\vec{r}}T_{00} (\vec{\xi}+\vec{r}) dr d\Omega= -\int_{A} \nabla \beta (r) \cdot \nabla_{\vec{r}}T_{00} (\vec{\xi}+\vec{r}) dr d\Omega
\end{equation}
%In the last equality, the surface term arising from green's identity vanishes because $\beta$ vanishes on $\p A$.% 
For $d=3$, \cite{Bhattacharya:2013fk} recently derived the same equation holographically.  In{\cite{Lashkari:2013uq}, a general argument was proposed explaining why \eqref{REE} leads to the perturbative Einstein's equations via the holographic entanglement entropy formula \eqref{RT}. 
In addition, a quantity called entanglement density was introduced in \cite{Nozaki:2013vta}. In $d=2$, for an interval \(A=[u,v] \) of length \(l=v-u\) and midpoint \(\xi\), this is defined as
\begin{align}
n(\xi, l, t) = \frac{1}{2} \frac{\d^{2} S_{A}}{\d u \d v}, \qquad 
\Delta n(\xi, l, t) = \frac{1}{2}\frac{\d^{2} \Delta S_{A}}{\d u \d v},
\end{align}
where in the second equality we present the shifted entanglement density in terms of the renormalized entanglement entropy \(\Delta S_{A} \).  Writing \(\Delta S_{A}\) in terms of $u$ and $v$ as in equation \eqref{HaR2} and computing the derivatives gives
\begin{align}
 l^{2} \Delta n(\xi, l, t) +  \Delta S_{A} =0,\\
 \lim_{l\R0}  \Delta n(\xi, l, t) =  T_{00}(\xi) \lim_{ l \R 0} 2\pi\int_{-l}^{l} dx' \frac{l^{2}-x'^{2}}{2l^{2}} = \frac{\pi}{3} T_{00}(\xi).
 \label{n}
 \end{align}
  which agrees with the holographic results of \cite{Nozaki:2013vta}.  
Finally, we note some overlap with \cite{bianchi}.
The author of \cite{bianchi} considered a gravitational theory on Rindler space and derived the change in entanglement entropy across the Rindler horizon as in equation (\ref{EE}) due to a metric perturbation \( g_{ab} \R g_{ab} +h_{ab} \).   There the entanglement Hamiltonian (\ref{K}), was evaluated along the event horizon H and was shown to be equal to an operator \(\hat{A}_{H}\) that measures the area of the event horizon. The crucial ingredient deriving this relation was the universal coupling \(\int h_{ab}T^{ab} \) of the graviton with the energy momentum tensor, which results in a perturbative Einstein's equation that relates \(T_{ab}\) to \(\Box h_{ab} \).  Thus, the renormalized horizon entanglement entropy was found to be
\begin{equation}
\d S_{H} = \frac{Tr (A_{H} \d \rho_{H})}{4G} =\frac{\d Area (H)}{4G}.
\end{equation}
Even though this equation was not derived from AdS/CFT, there is an obvious parallel here with  equation (\ref{dA}), where the minimal surface \(\gamma_{A}\) is identified with the horizon H.

%%%%%%%%%%%%%%%%%%%%%%%%%%%%%%%%%%%%%%%%%%%%%%%%%%%%%%%%%%%%%%%
   \section{Conclusion}
In this paper, we employed path integral methods to find a universal relation between the ground state entanglement Hamiltonian for an arbitrary region $A$ and the physical stress tensor.  For spherical entangling surfaces in a CFT we find, as in  \cite{myers}, that the entanglement Hamiltonian is the integral of a local density against a local entanglement temperature.  We further generalize this result to include states with conserved charges preserving conformal invariance and derive new expressions for the entanglement Hamiltonians in various cylindrical backgrounds in 2 dimensions.  Along the way, we show that the standard results for entanglement entropy in $d=2$ dimensions that are traditionally derived from the replica trick can be obtained easily by evaluating the \emph{thermal} entropy density using the entanglement temperature, and integrating over $A$.  While completing this paper, we became aware that the same method was used in  \cite{Swingle:2013hga} to obtain the leading area law behavior of entanglement entropy for a half space $A$ in a $d+1$-dimensional CFT and to derive the exact result for a finite interval $A$ in a $d=2$ CFT on the plane.    It was also argued there that at high temperatures the entanglement entropy for theories with a mass gap $m$ can be estimated by cutting off the size of the integration region A at \(x^{1}=\frac{1}{m}\), and indeed this gives the exact result for $d=2$.   In this paper, we made the additional observation that the entanglement temperature relates  the change in entanglement entropy to changes in conserved charges of the ground state via equation  (\ref{gflaw}).   However, we should note that the spatially varying entanglement temperature is not physical in the sense that it does not determine the expectation value of local observables such as \(T_{00}\) (Indeed, \(\la T_{00} \ra \) is a constant.). This is because the entanglement Hamiltonian (\ref{Ha}) is an integral over operators that do not commute, so the reduced density matrix does not factorize.   Indeed the entanglement temperature is not even conformally invariant; however equation \ref{gflaw} shows that in a fixed conformal frame, it gives a universal relation between the expectation value of physical charges  inside a region A and the  renormalized entanglement entropy.    

The relation (\ref{Ha}) between the entanglement Hamiltonian and the stress tensor, when combined with our CFT expression (\ref{EE}) for renormalized entanglement entropy provides a direct connection between the expectation value of the stress tensor and the increase in entanglement entropy, as was first noted in the holographic calculations of \cite{tak} and further generalized in \cite{Nozaki:2013vta}.   We also want to point out that it was recently observed in \cite{bha} that for spherical and cylindrical regions A, the holographic prescription \cite{rt} for the ground state entanglement entropy coincide with setting the \emph{finite} part of 00 component of the holographic stress tensor to zero on a 4D slice of the bulk spacetime.  The idea is that given a parametrization \(r=f(z)\) of the 4D slice, setting 
\begin{equation}\label{hst}
  \la T^{h}_{00} \ra = \la K_{ab}-h_{ab}K \ra =0,
   \end{equation} 
 where \(h_{ab}\)  is the induced metric and  \(K_{ab}\) is the extrinsic curvature 
gives a differential equation for \(f(z)\) that is identical to the minimal area equation.    A heuristic field theory justification might go as follows.  Demanding that a state in a CFT  has the same entanglement entropy of the ground state corresponds to setting 
\begin{equation}
\d S_{A}= \int_{A} \beta_{A} \la T^{CFT}_{00} \ra=0.
\end{equation} 
If we follow \cite{myers} we identify the region casual development of $A$  with a curved 4D slice of  AdS, then identifying \(\la T_{00}^{CFT} \ra \) with \(\la T^{h}_{00} \ra\) would gives equation \ref{hst}.

\section*{Acknowledgments}
We thank Phil Szepietowski for participation in the initial stages of this project. G.W. would like to thank Peter Arnold, Thomas Mark, and Yifei Shi for helpful discussions and the Michigan Center for Theoretical Physics for hospitality. L.PZ. would like to acknowledge the hospitality of University of Virginia through the Physics Theory Visitor program and University of Texas, Austin.
This work was supported by the DoE Grant \#DE-SC0007859 to the University of Michigan. G.W. was supported by the Presidential Fellowship from the University of Virginia. IK would like to acknowledge financial support from NSF CAREER award No. DMR-0956053. DV and GW were supported in part by DOE grant \#DE-SC0007984.

\appendix

%%%%%%%%%%%%%%%%%%%%%%%%%%%%%%%%%%%%%%%%%%%%%%%%%%%%%%%%%%%%%%%%%%%%%%%%%%%
\section{Evaluating the ground state entanglement entropy from the entanglement Hamiltonian}
%%%%%%%%%%%%%%%%%%%%%%%%%%%%%%%%%%%%%%%%%%%%%%%%%%%%%%%%%%%%%%%%%%%%%%%%%%%%%%
In this section we would like to point out a subtlety in evaluating the ground state entanglement entropy directly from equation (\ref{Ha}). The discussion will also serve to provide some background for the calculation in appendix \ref{App:NonU}.
Given the normalized reduced density, \(\rho_{A} =\exp(-H_{A})/Z_{A} \), the entanglement entropy is
\begin{equation}
S_{A}= -Tr_A(\rho_{A} \ln \rho_{A}) = Tr_A(\rho_{A} H_{A} )+ ln Z_{A}.
\end{equation}
Equation (\ref{Ha}) implies the entanglement energy vanishes:
\begin{equation}
\label{trH}
 Tr(\rho_{A} H_{A} ) = \int_A \beta(x) Tr(\rho T_{00})= \int_A \beta(x) \langle 0|:T_{00}:|0\rangle=0.
\end{equation}
In the last equality, we have normal ordered \(T_{00}\) with respect to the usual Minkowski annihilation operators, so \(S_{A}\) comes entirely from the "free energy " term\footnote{Even though \(Z_{A}\) is not an operator, we use the normal ordering symbol to highlight the fact that its  value depends on normal ordering. } \(ln :Z_{A}: \).
 
However there is an alternative way to evaluate the entanglement energy by conformally mapping \(\rho_{A}\) to a thermal density matrix with uniform temperature \cite{myers}.  In the case of a  a free scalar field in 2 spacetime dimensions and for \( A =\{ x > 0\}\),  \(H_{A}= 2\pi \int_{x\geq 0} x [(\partial_{x} \phi)^{2} +  (\partial_{t} \phi)^{2}]dx \) is the Rindler Hamiltonian \cite{susskind}.  In terms of Rindler coordinates
\begin{equation}
x=e^{\xi} \cosh(\eta), \qquad 
t=e^{\xi} \sinh(\eta),
\end{equation}
 it can be written as
\begin{align}
H_{A}= 2\pi \int_{-\infty}^{\infty}  (\partial_{\xi} \phi)^{2} +  (\partial_{\eta} \phi)^{2}d\xi. 
 \end{align}
 Thus \(H_{A}\) can be quantized by expanding the field in terms of plane waves in \emph{Rindler} coordinates \cite{unruh1984},
 \begin{align}\label{R}
 \phi_{k} = \int \frac{dk}{\sqrt{4\pi k}} b_{k} e^{ik(\xi - \eta)} +c.c. , \qquad 
  H_{A} = \int dk [b_{k}^{\dag}b_{k} + (1/2) \delta(0)] k.
 \end{align}
The delta function term represents the Casimir energy and is removed by normal ordering with respect to the \emph{Rindler} annihilation operator \(b_{k}\). It is well known  that under Rindler normal ordering, the Minkowski vacuum is thermal \cite {unruh1984} so that
 \begin{align}\label{RE}
Tr[\rho_{A}H_{A}]= \langle0|\vdots H_{A} \vdots|0\rangle = \frac{1}{12} \ln \frac{L}{\delta}
 \end{align}
 where $L$ and \(\delta\) are IR and UV cutoff's so that \(A=[\delta,L]\). This result can be obtained by a standard computation of the average thermal energy for a free relativistic gas of massless (Rindler) particles at temperature $\frac{1}{2\pi}$, subject to Dirichlet boundary conditions in  the \emph{Rindler} spatial coordinate  $\xi$.    Note that this differs from equation  (\ref{trH})  due to the difference in Rindler mode vs. Minkowski mode normal  ordering, which we denote by 3 and 2 dots respectively.    We can also obtain the corresponding Rindler free energy by usual statistical mechanics arguments:
 \begin{equation}\label{FE}
 \vdots \ln Z_{A} \vdots = \frac{1}{12} \ln \frac{L}{\delta}.
 \end{equation}
Adding this term to the entanglement energy (\ref{RE}) gives 
\begin{equation}
S_{A}= \frac{1}{6} \ln \frac{L}{\delta},
\end{equation} which is consistent with the known result \cite{wl}.    Since adding a normal ordering constant $a$ to \(H_{A}\) corresponds to a shift \(ln Z_{A} \rightarrow ln Z_{A} -a\), \eqref{FE}, \eqref{RE} and \eqref{trH} implies  $ :\ln Z_{A}: =\frac{1}{6} \ln \frac{L}{\delta} $, which is the same as $S_{A}$ as it should be.

The lesson here is that while $S_{A}$ is conformally invariant, neither the  entanglement energy or free energy is. 
 
To drive home this point we can derive the same result in a two dimensional Euclidean CFT, in the same spirit as \cite{myers} and \cite{hlw}.   The Euclidean version of the coordinate change from Minkowski to Rindler coordinates is the conformal map
 \begin{align}
 w= log z, \qquad  z=x+it, \qquad 
 w=\xi +i \theta,
 \end{align}
 where $z$ and $w$ are the  Euclideanised Minkowski/Rindler coordinates respectively and $ \theta$  is the angular coordinate on the $z$ plane. The $z$ plane is mapped to a strip of length $2 \pi$ and the Entanglement hamiltonian on the $z$ plane is mapped to physical hamiltonian \(H_{\theta}\) that evolves states along the \(\theta\) direction\footnote{This is the d=2 dimensional analogue of the conformal transformation to the hyperbolic space \(\mathcal{H}\) \ref{hung} for the half infinite line $A$.}.
For the ground state on the plane,  \(T(z)=0\)  so that the transformation of the stress tensor \footnote { To conform with the conventions of  \cite{cardy1988} $T_{ab}$ is defined so the Hamiltonian is $H= \frac{1}{2\pi} T_{00}$. } gives \(T(w)= c/24\) \cite{cardy1988} .  Integrating along \(\xi\)  to gives the expectation value of the  Hamiltonian  on the $z$ plane:
\begin{equation}
\langle  H_{\theta} \rangle = 2\pi  \int_{w(A)} d\xi  \frac{(\langle T(w)+T(\bar{w})\rangle)}{2\pi}= \frac{c}{12} \ln \frac{L}{\delta}
\end{equation}
which is the desired result\footnote{One of the 2\( \pi\)'s are from the length of the strip and the other from the definition of $H$ in terms of $T(w)$.}. In the last equality we have again set \(A=[\delta,L]\) on the $t =0$ slice of the $z$ plane, so that it is mapped to $w(A) =[\ln \delta, \ln (L)]$.

%%%%%%%%%%%%%%%%%%%%%%%%%%%%%%%%%%%%%%%%%%%%%%%%%%%%%%%%%%%%%%%%%%%%%%%%%%
\section{Non-uniform excitation of 2D free scalar field}\label{App:NonU}
In this appendix we  provide an explicit evaluation of the entanglement energy \(\d S_{A}= Tr(\delta \rho_{A} H_{A})\)  in equation  (\ref{EE1}) for a spatially non-uniform excitation   of a 2D free scalar field
and show that it is indeed equal to eq (\ref{EE}).  First note that normal ordering is irrelevant in this case because shifting \(H_{A}\) by a constant does not change the entanglement energy due to the normalization condition \(Tr (\d \rho_{A})=0 \).  Now following \cite{ben} we consider a particular excitation labelled by a positive Rindler momentum $k$:
 \begin{align}\label{d}
 d_{k}^{\dag} |0\rangle = \int_{0}^{\infty} dp D(k,p) a_{p}^{\dag} |0\rangle, \qquad D(k,p) = (2 k Sinh(\pi k))^{1/2}  \Gamma(-i k) |p|^{i k - \frac{1}{2}}, \qquad k>0
 \end{align}
where  \(a_{p}^{\dag}\) are the conventional Minkowski creation operators.  It is then straight forward to compute the (unnormalized) energy density by quantizing the energy density \(T_{00}= \frac{1}{2} \{(\partial_{\xi} \phi)^{2} +  (\partial_{\eta} \phi)^{2} \} \) in terms of Minkowski modes:
\begin{equation}
\langle 0| d_{k}: T_{00} : d^{\dag}_{k}|0\rangle = \frac{(-1+e^{2 k \pi}) k \pi^{2} {\rm csch}^{2}(k\pi)}{2\pi x^{2}}.
\end{equation}
Now we compute the \(\delta S_{A} \) for the half space A using the entanglement Hamiltonian in equation (\ref{Rindler}).
Dividing by the (infinite) normalization constant \(N=\langle 0| d_{k} d^{\dag}_{k}|0\rangle =2 \pi \int_{0}^{\infty} \frac{dp}{p} \) and inserting into the equation (\ref{EE1}) gives\footnote{The logarithmically divergent integral over x is cancelled by the normalization $N$.}
\begin{equation} \label{A}
\delta S_{A} = \pi k (1+\coth( k\pi)).
\end{equation}

Alternatively, we can evaluate \(\delta S_{A} \) using equation (\ref{EE}) via an explicit representation of the reduced density matrix \(\d \rho_{A} \) corresponding to the state in equation  (\ref{d}).  If we define the following reduced density matrices for the kth mode,
\begin{align}
\rho_{0}(k) = \sum_{n_{k}=0}^{\infty} e^{-2\pi n k} (1-e^{2 \pi k})|n_{k}\rangle \langle n_{k} |,\\
\rho_{1}(k)= \sum_{n_{k}=0}^{\infty} ( 4 n_{k} \sinh^{2} (\pi k)) |n_{k}\rangle \langle n_{k} |,
\end{align}
where \(|n_{k}\rangle\) denotes the occupational number basis for the Rindler particles, then results of  \cite{ben} imply that
\(\d \rho_{A} =  \rho_{1}(k)\prod_{l \neq k} \rho_{0}(l)- \prod_{l} \rho_{0}(l) \). Inserting this into (\ref{EE1}) and evaluating the trace using the Rindler Hamiltonian (\ref{R})  gives 
\begin{align}
\d S_{ A}  = 2 \pi \sum_{n\geq1} e^ { -2\pi n k} (4 n \sinh^{2}(\pi k)-   (1 - e^{-2 \pi k} ) \la n_{k}|\vdots H_{A} \vdots | n_{k}\rangle \\ \nonumber 
=  2 \pi \sum_{n\geq1} e^ { -2\pi n k} (4 n \sinh^{2}(\pi k)-   (1 - e^{-2 \pi k} ) ) n k \\
= \pi k (1+ \coth( \pi k) ), \nonumber 
\end{align}
 which is the same  result  as equation (\ref{A}).

\bibliographystyle{JHEP}
\bibliography{entanglement}

\end{document}